\input{epsf}

\documentstyle[11pt]{article}

\newtheorem{TT}{Theorem}
\newtheorem{CC}{Corollary}

\begin{document}
\baselineskip 4.0ex

\title{
{\Large\bf
Understanding Turbo Codes: A Signal Processing Study
}}

\author{
Xiang-Gen Xia\thanks{
Department of Electrical and Computer Engineering, University
of Delaware, Newark, DE 19716. 
Phone/Fax: (302)831-8038/4316. Email: xxia@ee.udel.edu. 
This work was supported in part 
by an initiative grant from the Department of Electrical
and Computer Engineering, University of Delaware,
 the Air Force Office of Scientific Research 
(AFOSR) under Grant No. F49620-97-1-0253, and 
the National Science Foundation CAREER Program under Grant
MIP-9703377.}
}

\date{June 22, 1997}

\maketitle

%

\begin{abstract}
In this paper, we study turbo codes from the digital
signal processing point of view by defining turbo codes
over the complex field. It is known that iterative
decoding and interleaving
between concatenated parallel codes are two key elements that make
turbo codes perform significantly better than the conventional
error control codes. This is analytically illustrated in this paper by showing
that the decoded noise mean power in the iterative decoding  decreases
when the number of iterations increases,
as long as  the interleaving decorrelates
the noise after each iterative decoding step. 
An analytic decreasing rate and the limit 
of the decoded noise mean power are given. 
The limit of the decoded noise mean power of the iterative 
decoding of a turbo code with two parallel
codes with their rates less than $1/2$ is one third 
of the noise power before the decoding,
which can not be achieved by any non-turbo codes
with the same rate.
From this study, the role of designing a good interleaver
can also be clearly seen.

\end{abstract}

\section{Introduction}
\setcounter{equation}{0}

Turbo codes have recently received considerable attentions, see
for example [1-9],   due to the near Shannon limit
performance. The recent research results [2]
on turbo codes have indicated that the bit error rate (BER)
 can reach $10^{-5}$ at 
signal-to-noise ratio (SNR) $E_b/N_0=0.0$dB. This ``magic'' performance
has been surprising the communication community lately, and
meanwhile questions, doubts, and explanations have been followed up too.
The main question is: what is the magic of turbo codes? 
In turbo codes, there are two key elements: iterative decoding
and interleaving between concatenated parallel codes. 
A simple example of rate 1/2 
turbo code is shown in Fig. 1(a) for encoding and
Fig. 1(b) for decoding, where $\downarrow 2$ means 
downsampling by $2$ , i.e., taking one in each two samples,
and $z^{-1}$ denotes the delay variable as denoted by $D$
in the channel coding literature.
An intuitive explanation for the super
performance of turbo codes is the following. 

Because of the hardware limitations, the error correction
capability in low SNR environments is  limited due
to the limited distance property of a limited size error correction code.
This means that implementing a decoding algorithm once may not be able to
correct all errors  but some errors, i.e., the total
number of errors is reduced. A natural question is, to correct more errors,
why one does not implement the decoding
algorithm for the decoded signal again. To answer this 
question, one should notice that, although the total number 
of errors after the first decoding is reduced, the decoding algorithm
may cause patterned errors, such as burst errors. 
These patterned errors may increase the number of total
errors of the second implementation of the decoding algorithm. 
This implies that 
 the direct implementation of cascaded multiple decoding algorithms 
may not be helpful. To get around of the patterned errors,
an intuitive method is to add an interleaver between 
two cascaded decoding algorithms to
spread  the patterned errors over a long period of time.
The interleaving between two decoders forces 
the interleaving  between the concatenated encoders.

\begin{figure}[htbp]
 \begin{center}
 \epsfysize=5.0in
 \leavevmode\epsffile[8 180 581 707]{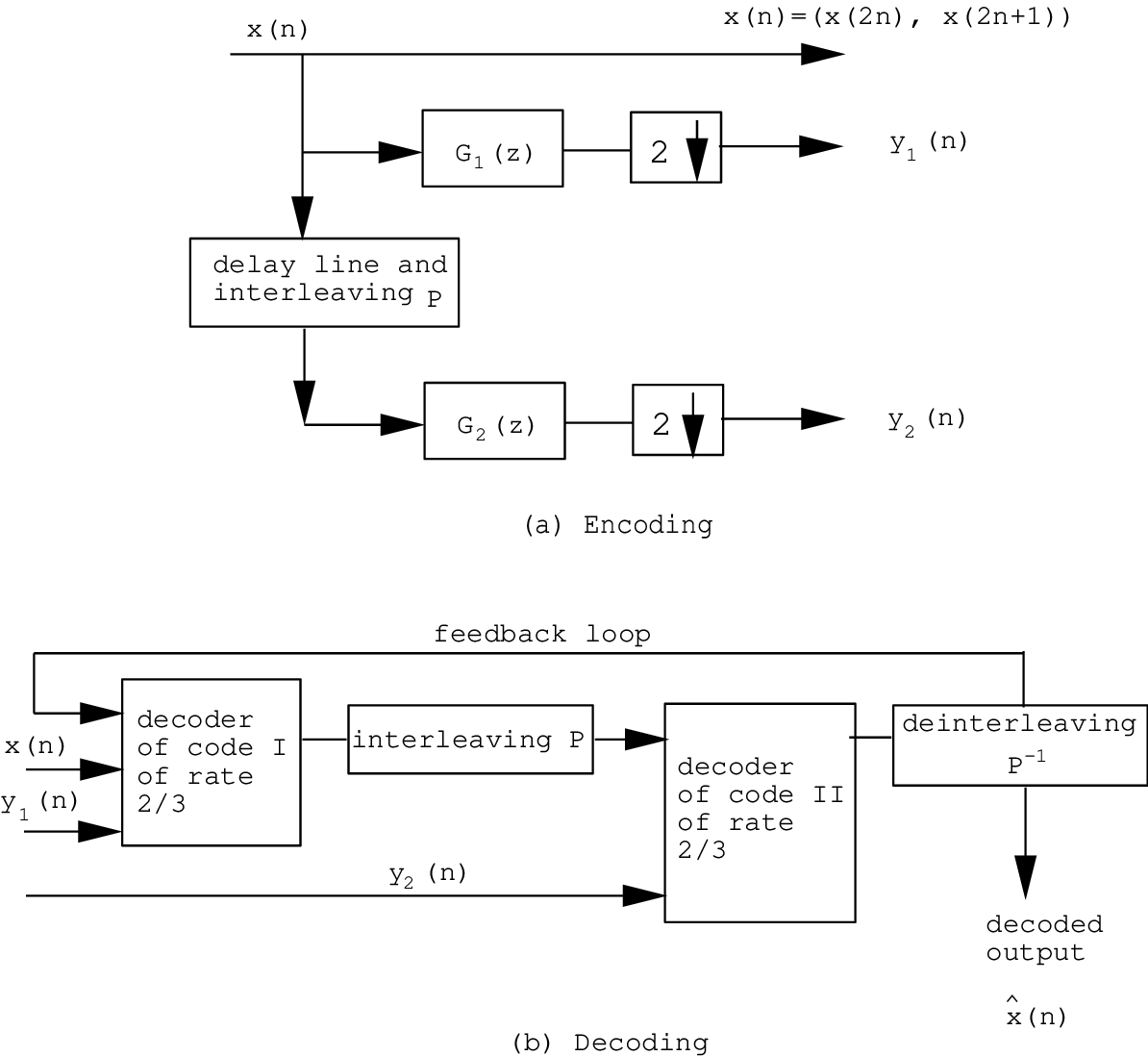}
\end{center}
\caption{A rate 1/2 turbo code.}
\end{figure}

Although the above intuitive explanation is sometimes
good enough, a quantitative analysis is more important
to understand turbo codes better. 
There have been several excellent 
articles in the literature on the performance
analysis of turbo codes, such as [3-9]. 
A strong analytic analysis is still needed
to show the better performance of turbo codes
than the conventional codes. Most recently,
Forney in [18] mentioned ``turbo codes
work very well, but we don't understand them.''
The goal of this paper is to study turbo codes
from the digital signal processing perspective
by extending  turbo codes and other codes defined from finite
fields to the  the complex field, and comparing 
them on the complex field. 
 Since all codes here are defined over 
the complex field, the maximal likelihood decoding
is the least squared error decoding, which 
can be achieved by solving a system of linear equations.
By doing so, we show that the decoded noise mean power decreases
when the iteration number in the decoding  increases
under the condition that the interleaving
between concatenated parallel codes decorrelates 
the noise after each decoding iteration. Given a rate of
a turbo code, the decreasing rate of the decoded noise mean power
in terms of the number of iterations in the decoding 
is given, and the limit of the decoded noise mean power is also given 
when the number of iterations tends to infinity. 
It turns out that the decoded noise mean power in the limit
sense can not be achieved by any non-iterative
codes, i.e., non-turbo codes, with the same 
rate. 

This paper is organized as follows. In Section 2,
we study a general systematic error control code
defined on the complex field and formulate
the noise power after the least squared error decoding.
The study in this section provides a foundation for the late 
analysis. 
In Section 3, we study the decoded noise mean power after the iterative turbo 
code decoding. In Section 4, we present some numerical
examples.

\section{Linear Error Control Codes Over the Complex Field}
\setcounter{equation}{0}

To study turbo codes defined on the complex field, 
in this section we first study a general linear error control code
defined over the complex field, which have been recently
discussed in [10]. As mentioned in [10], there are 
two advantages of error control codes defined on the complex
field over the conventional error control codes. 
Since error control codes are defined on the complex field,
all arithmetics are in the complex field and therefore the
maximal likelihood decoding is the least squared error
solution, that is the solution of a linear system. 
This is the first advantage, i.e., linear algorithm decoding.
Since all channel distortions are with the complex field
arithmetic,
error control codes over the complex field can be designed
to completely cancel an FIR intersymbol interference (ISI),
which is not possible for any error control codes defined
on finite fields. In [11-13], channel independent 
error control codes are designed to cancel the  ISI,
which are called {\em ambiguity resistant codes}. 
This is the second advantage. 

In this section, we focus on channels with only additive 
random noise and study the noise power change after the least
squared error decoding.

A linear error control code defined on the complex
field is usually represented by
\begin{equation}\label{2.1}
Y(z)=G(z)X(z),
\end{equation}
where $X(z)=\sum_n X(n)z^{-n}$, $G(z)=\sum_{n=0}^{Q_G} G(n)z^{-n}$,
and $Y(z)=\sum_n Y(n)z^{-n}$ are the $z$-transform of $K\times 1$
input vector sequence $X(n)$, $N\times K$ code generator 
matrix with order $Q_G$, and 
the $z$-transform of $N\times 1$
output vector sequence $Y(n)$, respectively. In particular,
for a rate $K/N$ block linear error control code,
\begin{equation}\label{2.2}
Y=GX,
\end{equation}
where $X$, $Y$ and $G$ are $K\times 1$, $N\times 1$ and $N\times K$
(vectors) matrices, respectively. On the other hand, a convolutional
code in (\ref{2.1}) can  also be represented by the block representation
(\ref{2.2}), where $G$ is the generalized Sylvester matrix of $G(z)$:
\begin{equation}\label{2.3}
\left[ \begin{array}{ccccc}
G(Q_G) &  \cdots & G(0)&  \cdots& 0 \\
\vdots & \ddots & \ddots & \ddots &  \vdots \\
0 & \cdots  & G(Q_G) &  \cdots & G(0)
\end{array} \right ].
\end{equation}
Because of the mutual representations of block and convolutional
codes in (\ref{2.1})-(\ref{2.3}), 
without loss of generality, we focus on the block representation
(\ref{2.2}). A systematic rate $K/N$ block code $G$ is
\begin{equation}\label{2.4}
G=\left[ \begin{array}{c}I_K \\ F \end{array}\right],
\end{equation}
where $I_K$ is the $K\times K$ identity matrix and 
$F$ is an $(N-K)\times K$ constant matrix.
Since, in turbo codes, systematic codes are usually used, in the rest
of this section we focus on systematic block codes. Let $G$ in (\ref{2.4})
be
\begin{equation}\label{2.5}
G=\left[ \begin{array}{c}I_K \\ F \end{array}\right]=
\left[ \begin{array}{cccc}
1 & 0 & \cdots & 0 \\
0 & 1 & \cdots & 0\\
\vdots & \vdots & \vdots & \vdots \\
0 & 0 & \cdots & 1\\
g_{K+1,1} & g_{K+1,2} & \cdots & g_{K+1,K}\\
\vdots & \vdots & \vdots & \vdots \\
g_{N,1} & g_{N,2} & \cdots & g_{N,K}
\end{array} \right].
\end{equation}
To preserve the code output signal power, i.e.,
the output signal power is equal to the input signal power,
 the following
normalization on $G$ is imposed:
\begin{equation}\label{2.6}
\sum_{k=1}^K |g_{n,k}|^2=1,\,\,K+1\leq n\leq N.
\end{equation}
The received signal is
\begin{equation}\label{2.7}
\tilde{Y}=GX+\eta,
\end{equation}
where
$X=(X(1),\cdots, X(K))^T$,
 $\tilde{Y}=(\tilde{Y}(1),\cdots, \tilde{Y}(N))^T$,
$\eta=(\eta(1),\cdots, \eta(N))^T$ 
are the input signal, the received signal and the additive random noise,
respectively, and  $^T$ means the transpose. The additive noise
$\eta(n)$ is assumed i.i.d. and with mean $0$ and variance (or power)
$\sigma_{\eta}^2$.

The maximal likelihood decoding of (\ref{2.7}) is the least squared
error solution:
\begin{equation}\label{2.8}
\hat{X}=(G^{\dagger}G)^{-1}G^{\dagger}
\tilde{Y}=X+(G^{\dagger}G)^{-1}G^{\dagger}\eta,
\end{equation}
where $^{\dagger}$ means the transpose conjugate. Thus,
the noise after the decoding is
\begin{equation}\label{2.9}
\hat{\eta}=(G^{\dagger}G)^{-1}G^{\dagger}\eta,
\end{equation}
where $\eta$ is the additive noise before decoding and has power 
$\sigma_{\eta}^2$. We next want to calculate the mean power of this
decoded noise $\hat{\eta}$.

Let $\eta=(\eta_1^T, \eta_2^T)^T$. Then $\eta_1$ and $\eta_2$ are
independent each other. It is not hard to see that
\begin{equation}\label{2.10}
\hat{\eta}=(I_K+F^{\dagger}F)^{-1}(\eta_1+F^{\dagger}\eta_2).
\end{equation}
Let the singular value decomposition of $F$ be
\begin{equation}\label{2.11}
F=U\Lambda V,
\end{equation}
where $U=(u_{k,l})$ and $V=(v_{m,n})$ are $N\times N$ and $K\times K$ unitary matrices,
respectively, 
and $\Lambda$ is the diagonal matrix.
There are two cases for the diagonal matrix $\Lambda$.

\noindent
{\bf Case I:} $N-K<K$.

In this case, 
\begin{equation}\label{2.12}
\Lambda =\left[ 
\mbox{diag}(\alpha_1,\cdots,\alpha_{N-K}), 0_{(N-K)\times (2K-N)}
\right],
\end{equation}
where 
$\alpha_1$, ..., $\alpha_{N-K}$ with $|\alpha_1|\geq \cdots \geq |\alpha_{N-K}|$
are the singular values of $F$. Then (\ref{2.10})
becomes 
$$
\hat{\eta}=V^{\dagger}\mbox{diag}\left(
\frac{1}{1+|\alpha_1|^2},\cdots, \frac{1}{1+|\alpha_{N-K}|^2}, 1, \cdots, 1
\right)
(V\eta_1+\Lambda^{\dagger}U^{\dagger}\eta_2).
$$

Let 
$\hat{\eta}=(\hat{\eta}(1),\cdots, \hat{\eta}(K))^T$, then, by some algebra,
the power of the $k$th component $\hat{\eta}(k)$
of $\hat{\eta}$ is
\begin{equation}\label{2.13}
\sigma_{\hat{\eta}(k)}^2=
\sigma_{\eta_1}^2\left(\sum_{m=1}^{N-K} |v_{m,k}|^2\frac{1}{(1+|\alpha_m|^2)^2}
+\sum_{m=N-K+1}^{K} |v_{m,k}|^2\right)
+\sigma_{\eta_2}^2\sum_{m=1}^{N-K} |v_{m,k}|^2\frac{|\alpha_m|^2}{(1+|\alpha_m|^2)^2},
\end{equation}
where $v_{m,k}$ are the entries of the unitary matrix $V$ and
$\sigma_{\eta_i}^2$ is the power of $\eta_i$ for $i=1,2$,
which are equal to $\sigma_{\eta}^2$.
Therefore, by the unitariness of the matrix $V$,
  the mean power of the decoded noise  $\hat{\eta}$ is
\begin{equation}\label{2.14}
\sigma_{\hat{\eta}}^2
=\frac{\sigma_{\eta_1}^2}{K}\left( 
\sum_{k=1}^{N-K}\frac{1}{(1+|\alpha_k|^2)^2} 
+2K-N \right)+\frac{\sigma_{\eta_2}^2}{K} 
\sum_{k=1}^{N-K}\frac{|\alpha_k|^2}{(1+|\alpha_k|^2)^2}.
\end{equation}
Since $\sigma_{\eta_1}^2=\sigma_{\eta_2}^2=\sigma_{\eta}^2$,
\begin{equation}\label{2.15}
\sigma_{\hat{\eta}}^2=
\frac{\sigma_{\eta}^2}{K} \left(
\sum_{k=1}^{N-K}\frac{1}{1+|\alpha_k|^2}+2K-N\right).
\end{equation}

By the normalization (\ref{2.6}),
\begin{equation}\label{2.16}
\sum_{k=1}^{K}|\alpha_k|^2=\sum_{n=K+1}^N \sum_{k=1}^K |g_{n,k}|^2=N-K,
\end{equation}
where $\alpha_{N-K+1}=\cdots =\alpha_K=0$.

Given a rate $K/N$, by (\ref{2.16}) 
the above mean power $\sigma_{\hat{\eta}}^2$ in (\ref{2.15})
reaches the minimum {\em when and only when}
$|\alpha_k|=1$ for $1\leq k\leq N-K$ and the minimum is
\begin{equation}\label{2.17}
\sigma_{\hat{\eta}}^2=\frac{\sigma_{\eta}^2}{K}
\left(\frac{N-K}{2}+2K-N\right)= 
\left( 1-
\frac{N-K}{2K}\right)\sigma_{\eta}^2.
\end{equation}
In this case, the above minimum mean power can be expressed
in terms of the powers $\sigma_{\eta_i}^2$ for $i=1,2$ as
\begin{equation}\label{2.18}
\sigma_{\hat{\eta}}^2
= \frac{7K-3N}{4K}\sigma_{\eta_1}^2+
\frac{N-K}{4K}\sigma_{\eta_2}^2.
\end{equation}
This formula will be used in Section 3 later.

\noindent
{\bf Case II:} $N-K\geq K$.

In this case, 
\begin{equation}\label{2.19}
\Lambda =\left[ \begin{array}{c}
\mbox{diag}(\alpha_1,\cdots,\alpha_{K})\\
0_{(N-K)\times K} \end{array} \right],
\end{equation}
where 
$\alpha_1$, ..., $\alpha_{K}$ with $|\alpha_1|\geq \cdots \geq |\alpha_{K}|$
are the singular values of $F$. Then (\ref{2.10})
becomes 
$$
\hat{\eta}=V^{\dagger}\mbox{diag}\left(
\frac{1}{1+|\alpha_1|^2},\cdots, \frac{1}{1+|\alpha_K|^2}\right)
(V\eta_1+\Lambda^{\dagger}U^{\dagger}\eta_2).
$$

Similar to (\ref{2.13}), 
the power of the $k$th component $\hat{\eta}(k)$
of $\hat{\eta}$ is
\begin{equation}\label{2.20}
\sigma_{\hat{\eta}(k)}^2=
\sigma_{\eta_1}^2 \sum_{m=1}^{K} |v_{m,k}|^2\frac{1}{(1+|\alpha_m|^2)^2}
+\sigma_{\eta_2}^2\sum_{m=1}^{K} |v_{m,k}|^2\frac{|\alpha_m|^2}{(1+|\alpha_m|^2)^2}.
\end{equation}
Thus, the mean power of $\hat{\eta}$ is
\begin{equation}\label{2.21}
\sigma_{\hat{\eta}}^2=
\frac{\sigma_{\eta_1}^2}{K}
 \sum_{k=1}^{K}\frac{1}{(1+|\alpha_k|^2)^2}
+\frac{\sigma_{\eta_2}^2}{K}\sum_{k=1}^{K} \frac{|\alpha_k|^2}{(1+|\alpha_k|^2)^2}=
\frac{\sigma_{\eta}^2}{K}
\sum_{k=1}^{K} \frac{1}{1+|\alpha_k|^2}.
\end{equation}

Given a rate $K/N$, by (\ref{2.16})
the above mean power $\sigma_{\hat{\eta}}^2$
reaches the minimum {\em when and only when}
$|\alpha_k|^2=(N-K)/K$ for $1\leq k\leq K$ and the minimum is
\begin{equation}\label{2.22}
\sigma_{\hat{\eta}}^2=\frac{\sigma_{\eta}^2}{K}
\frac{K}{1+(N-K)/K}=\frac{K}{N}\sigma_{\eta}^2.
\end{equation}
In terms of $\sigma_{\eta_i}^2$ for $i=1,2$, the minimum
decoded noise mean power is 
\begin{equation}\label{2.23} 
\sigma_{\hat{\eta}}^2=\frac{K^2}{N^2}\sigma_{\eta_1}^2
+\frac{K(N-K)}{N^2}\sigma_{\eta_2}^2. 
\end{equation}

This concludes the following theorem.

\begin{TT}
Let a rate $K/N$ systematic error control code $G$ have 
the form (\ref{2.5})-(\ref{2.6}). The decoded noise 
mean power 
$\sigma_{\hat{\eta}}^2$ after the least squared error
decoding is expressed in  (\ref{2.15}) when $N-K<N$,
and  in (\ref{2.21}) when $N-K\geq N$, where $\alpha_k$
are the singular values of the $(N-K)\times K$ matrix $F$ in code $G$.
The decoded noise mean power $\sigma_{\hat{\eta}}^2$
reaches the minimum if and only if 
\begin{equation}\label{2.24}
|\alpha_k|^2=\left\{ \begin{array}{ccc} 
1, & 1\leq k\leq N-K, &\mbox{when } N-K<K,\\
\frac{N-K}{K}, & 1\leq k\leq K, & \mbox{when } N-K\geq K.
\end{array}\right.
\end{equation}
Moreover, the minimum decoded noise mean power is
\begin{equation}\label{2.25}
\sigma_{\hat{\eta}}^2=\left\{ \begin{array}{cc} 
\left(1-\frac{N-K}{2K}\right) \sigma_{\eta}^2, &\mbox{when } N-K<K,\\
\frac{K}{N} \sigma_{\eta}^2, & \mbox{when } N-K\geq K,
\end{array}\right.
\end{equation}
where $\sigma_{\eta}^2$ is the noise power before
the decoding.
\end{TT}

This theorem leads to the following corollary.

\begin{CC}
A rate $K/N$ systematic code $G=(I_K, F^T)^T$ is optimal
if and only if  the $N-K$ nonzero eigenvalues of the matrix $F^{\dagger}F$
are equal to $1$ when $N-K<K$, and all eigenvalues 
of the matrix $F^{\dagger}F$
are equal to $(N-K)/K$ when $N-K\geq K$.
\end{CC}

From these results, one can see that the noise mean
power $\sigma_{\hat{\eta}}^2$ after the least squared error
decoding is less than the original noise power $\sigma_{\eta}^2$,
which gives us the following {\em coding gain}
for the optimal rate $K/N$ systematic codes over the complex field:
\begin{equation}\label{2.26}
\gamma (K,N) \stackrel{\Delta}{=}
\left\{ \begin{array}{cc}
\frac{2K}{2K-(N-K)}, & \mbox{when } N-K<K,\\
\frac{N}{K}, & \mbox{when } N-K\geq K.
\end{array} \right.
\end{equation}
Clearly, the coding gain $\gamma (K,N)>1$ when $N>K$. 
Then the decoded noise mean power can be expressed in terms of
the original noise mean power and the coding gain:
\begin{equation}\label{2.27}
\sigma_{\hat{\eta}}^2=\frac{1}{\gamma(K,N)}\sigma_{\eta}^2,
\end{equation}
i.e., the decoded noise mean power is reduced by the factor 
of $1/\gamma(K,N)$. 

Notice that the decoded noise mean power 
$\sigma_{\hat{\eta}}^2$ is calculated from the LSE
solution formula (\ref{2.10}). The {\em key}
for this power $\sigma_{\hat{\eta}}^2$ less than
the original noise power $\sigma_{\eta}^2$ is the
independence of the two noise parts $\eta_1$ and $\eta_2$
in (\ref{2.10}).  
The maximal noise power reduction is achieved when
$\eta_1$ and $\eta_2$ are completely independent.
The following simple example
shows that the noise power reduction property may not hold
when these two noise parts are correlated. 
Let $F=I_K$ and $\eta_1=\eta_2$ in (\ref{2.10}). 
Then $\hat{\eta}=\eta_1=\eta_2$. 
In the next section, the interleaving plays the role
of {\em decorrelating} these two noise parts generated
at each iterative decoding step in turbo code decoding,
which is the {\em key} for turbo codes to have a good performance as
we will see later.

Another remark we want to make here is the connection
between the above coding gain and the spreading gain
in CDMA communication 
systems, see for example [17]. In CDMA systems, each symbol 
is spread into $N$ symbols, where $N$ is the signature
length. This case corresponds to the above coding with
$K=1$. Clearly, the coding gain in this case is also $N$,
i.e., the coding gain and the spreading gain are the same.
The MMSE multiuser detection, [14-16], is similar to the above
LSE decoding.

\section{Turbo Codes Over the Complex Field}
\setcounter{equation}{0}

\begin{figure}[htbp]
 \begin{center}
 \epsfysize=5.6in
 \leavevmode\epsffile[8 119 581 706]{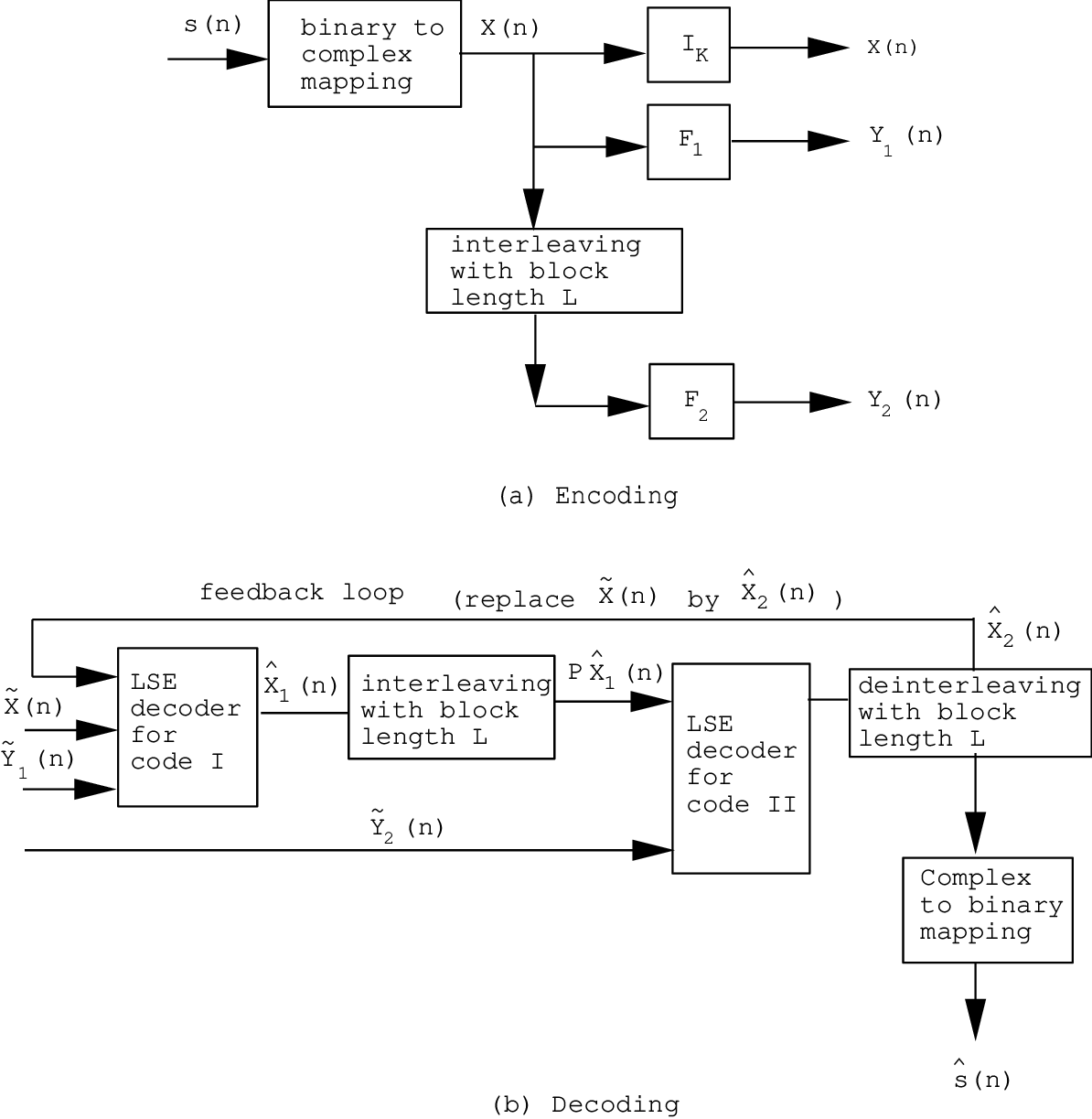}
\end{center}
\caption{A turbo code with two parallel codes defined on the complex field.}
\end{figure}

After the previous study of systematic codes on the complex
field, in this section we want to study turbo codes over the
complex field, in particular, the decoded noise mean power change
in the iterative decoding algorithm. Notice that, in the least
squared error (LSE) decoding studied in Section 2, no iteration
is used. We show that the iterative decoding makes 
the decoded noise mean power decrease when the number of iterations
increases as long as the interleaver between parallel decoders
decorrelates the decoded noise well after each decoder.

Similar to what was mentioned in Section 2, without loss of generality
we only consider linear block codes in turbo codes as parallel codes
with total {\em two} parallel codes as shown in Fig. 2.
To study the turbo code shown in Fig. 2, we first 
describe some notations and formulations.

The rate of the turbo code shown in Fig. 2 is $K/N$,
where the matrix $F_1$ has size $(N_1-K)\times K$, 
the matrix $F_2$ has size $(N_2-K)\times K$, and $N=N_1+N_2-K$. Let
\begin{equation}\label{3.1}
G_1=\left[ \begin{array}{c}I_K\\F_1 \end{array} \right],\,\,\mbox{ and }\,\,
G_2=\left[ \begin{array}{c}I_K\\F_2 \end{array} \right].
\end{equation}
Let $P_L$ denote the interleaver with block length $L$ in Fig. 2
and $P_L^{-1}$ be its inverse. Then the turbo code encoding in
Fig. 2(a) is
\begin{equation}\label{3.2}
\left[ \begin{array}{c}X(n)\\ Y_1(n) \end{array} \right]
=G_1X(n),\,\,\mbox{ and }\,\,
Y_2(n)=F_2(P_LX)(n),
\end{equation}
where $n$ is the time index and the interleaver
$P_L$ acts not only on the $K\times 1$ vector $X(n)$
at the time $n$
but also on a sequence of these vectors with total 
grouped vector length $L$. The received signal is 
\begin{equation}\label{3.3}
\left[ \begin{array}{c}
\tilde{X}(n)\\ \tilde{Y}_1(n) \\ \tilde{Y}_2(n) \end{array} \right]
= \left[ \begin{array}{c}  X(n)\\ Y_1(n) \\ Y_2(n) \end{array} \right]
+ \left[ \begin{array}{c}  \eta_0(n)\\ \eta_1(n) \\ \eta_2(n) \end{array} \right]= \left[ \begin{array}{c} X(n)+\eta_0(n)\\ Y_1(n)+\eta_1(n) \\ Y_2(n)+\eta_2(n) \end{array} \right],
\end{equation}
where $\eta_0(n),\eta_1(n), \eta_2(n)$ 
are all together i.i.d. with mean $0$ and variance (or power)
$\sigma_{\eta}^2$, i.e.,
\begin{equation}\label{3.4}
\mbox{E}\eta_i(n)=0,\,\,\,\,\,\mbox{E}|\eta_i(n)|^2=\sigma_{\eta}^2,
\,\,\, i=0,1,2.
\end{equation}

We next want to formulate the output noise after each iteration
of the decoding in Fig. 2(b) and then study the output noise
mean power. Notice that the LSE decoder for codes I and II in Fig. 2(b)
are the least squared error solutions for code $G_1$ and $G_2$ in
(\ref{3.1}) as studied in Section 2. Thus the first step
decoded signals 
in the iterative decoding in Fig. 2(b) are
\begin{eqnarray}
\hat{X}_1^{(1)}(n) & \stackrel{\Delta}{=} & \hat{X}_1(n)= (G_1^{\dagger}G_1)^{-1}G_1^{\dagger}
\left[ \begin{array}{c}  \tilde{X}(n)\\ \tilde{Y}_1(n)\end{array} \right],
\label{3.5}\\
\hat{X}_2^{(1)}(n) & \stackrel{\Delta}{=} & \hat{X}_2(n)= 
P_L^{-1}(G_2^{\dagger}G_2)^{-1}G_2^{\dagger}
\left[ \begin{array}{c} (P_L\hat{X}_1^{(1)})(n)\\ \tilde{Y}_2(n)\end{array} \right].
\label{3.6}
\end{eqnarray}
We will see later that the interleaver
$P_L$ in (\ref{3.6}) is for decorrelating the decoded
noise, which makes the interleaving in the encoding
necessary. 

Similar to (\ref{2.8})-(\ref{2.9}), the new 
noises after the LSE decoders for codes $G_1$ and $G_2$ at the first iteration
are
\begin{eqnarray}
\hat{\eta}_1^{(1)}(n) & \stackrel{\Delta}{=} & \hat{X}_1^{(1)}(n)
-X(n)= (I_K+F_1^{\dagger}F_1)^{-1} (\eta_0(n)+F_1^{\dagger}\eta_1(n)),
\label{3.7}\\
\hat{\eta}_2^{(1)}(n) & \stackrel{\Delta}{=} & \hat{X}_2^{(1)}(n)
-X(n)= P_L^{-1}(I_K+F_2^{\dagger}F_2)^{-1} (
P_L \hat{\eta}_1^{(1)}(n)+F_2^{\dagger}\eta_2(n)).
\label{3.8}
\end{eqnarray}
For the second iteration, 
$\hat{\eta}_2^{(1)}(n)$ is the new noise of the decoded signal
$\hat{X}_2(n)$ (replacing $\tilde{X}(n)$) in the new input signal.
For the formulas at the $l$th iteration, 
we first have the following notations.

Let 
\begin{equation}\label{3.9}
\hat{X}_i^{(0)}(n)\stackrel{\Delta}{=}\tilde{X}(n),\,\,\,\,i=1,2.
\end{equation}
Let  $\hat{X}_1^{(l)}(n)$
and $\hat{X}_2^{(l)}(n)$ 
be the outputs of the LSE decoders for codes $G_1$ and $G_2$
at the $l$th iteration, respectively. Let 
$\hat{\eta}_i^{(l)}(n)$ denote the new noise after the LSE
decoder for code $G_i$ at the $l$th iteration for $i=1,2$.
Then, from (\ref{3.7})-(\ref{3.8}), it is not hard to
see that the following general formulas for the new noises hold.

For $l=1,2,...$,
\begin{eqnarray}
\hat{\eta}_1^{(l)}(n) & \stackrel{\Delta}{=} & \hat{X}_1^{(l)}(n)
-X(n)= (I_K+F_1^{\dagger}F_1)^{-1} (\hat{\eta}_2^{(l-1)}(n)+F_1^{\dagger}\eta_1(n)),
\label{3.10}\\
\hat{\eta}_2^{(l-1)}(n) & \stackrel{\Delta}{=} & \hat{X}_2^{(l)}(n)
-X(n)= P_L^{-1}(I_K+F_2^{\dagger}F_2)^{-1} (
P_L \hat{\eta}_1^{(l-1)}(n)+F_2^{\dagger}\eta_2(n)).
\label{3.11}
\end{eqnarray}

If $\hat{\eta}_2^{(l-1)}(n)$ and $F_1^{\dagger}\eta_1(n)$ are independent
and $P_L\hat{\eta}_1^{(l-1)}(n)$ and $F_2^{\dagger}\eta_2(n)$ are independent,
then the mean powers of $\hat{\eta}_1^{(l)}(n)$ and 
$\hat{\eta}_2^{(l)}(n)$ in (\ref{3.10})-(\ref{3.11}) 
can be calculated similar to (\ref{2.18}) and (\ref{2.23})
for the decoded noise $\hat{\eta}$ in (\ref{2.10}). To study 
the independences, let us see the first two iterations in 
(\ref{3.7})-(\ref{3.11}) for these new noises.

Let 
\begin{equation}\label{3.12}
A\stackrel{\Delta}{=}(I_K+F_1^{\dagger}F_1)^{-1},\,\,\,\,
B\stackrel{\Delta}{=}(I_K+F_2^{\dagger}F_2)^{-1}.
\end{equation}
Then
\begin{eqnarray*}
\hat{\eta}_1^{(1)}(n) & = & A\eta_0(n)+AF_1^{\dagger}\eta_1(n),\\
\hat{\eta}_2^{(1)}(n) & = & P_L^{-1}BP_L A \eta_0(n)+
 P_L^{-1}BP_LAF_1^{\dagger}\eta_1(n)+
P_L^{-1}BF_2^{\dagger}\eta_2(n),\\
\hat{\eta}_1^{(2)}(n) & = & A\hat{\eta}_2^{(1)}(n)+AF_1^{\dagger}\eta_1(n)\\
                   & = & 
   AP_L^{-1}BP_L A \eta_0(n)
+ AP_L^{-1}BP_LAF_1^{\dagger}\eta_1(n)+ AF_1^{\dagger}\eta_1(n)
+AP_L^{-1}BF_2^{\dagger}\eta_2(n),\\
\hat{\eta}_2^{(2)}(n) & = & P_L^{-1}BP_L \hat{\eta}_1^{(2)}(n)+
                         P_L^{-1}BF_2^{\dagger}\eta_2(n)\\ 
                   & = &  
   P_L^{-1}BP_LAP_L^{-1}BP_L A \eta_0(n)+ 
 P_L^{-1}BP_LAP_L^{-1}BP_LAF_1^{\dagger}\eta_1(n)
 + P_L^{-1}BP_LAF_1^{\dagger}\eta_1(n)\\
  & &   +P_L^{-1}BP_LAP_L^{-1}BF_2^{\dagger}\eta_2(n) 
+P_L^{-1}B F_2^{\dagger}\eta_2(n).
\end{eqnarray*}

Let 
\begin{equation}\label{3.13}
Q\stackrel{\Delta}{=}P_L^{-1}BP_LA=P_L^{-1}(I+F_2^{\dagger}F_2)^{-1}P_L(I+F_1^{\dagger}F_1)^{-1}.
\end{equation}
Then the above expressions can be generalized to
\begin{eqnarray}
\hat{\eta}_1^{(l)}(n) & = & A\left(
Q^{l-1}\eta_0(n)+\sum_{i=0}^{l-1}Q^iF_1^{\dagger}\eta_1(n)
+\sum_{i=0}^{l-2}Q^iP_L^{-1}BF_2^{\dagger}\eta_2(n)\right),\label{3.14}\\
\hat{\eta}_2^{(l)}(n) & = & 
Q^{l}\eta_0(n)+\sum_{i=1}^{l}Q^iF_1^{\dagger}\eta_1(n)
+\sum_{i=0}^{l-1}Q^iP_L^{-1}BF_2^{\dagger}\eta_2(n),\label{3.15}
\end{eqnarray}
where $l=2,3,4,...$.

With the above formulations, we have proved the following theorem on
the independences.

\begin{TT}
Let $A$, $B$, $Q$ be the operators defined in (\ref{3.12})-(\ref{3.13}),
$F_1$ and $F_2$ be as in (\ref{3.1}), and 
$P_L$ be an interleaver with block length $L$.
If 
\begin{itemize}
\item[(a)] 
$Q^{k_1}F_1^{\dagger}\eta_1(n)$ and  $Q^{k_2}F_1^{\dagger}\eta_1(n)$ 
are independent for any $0\leq k_1\neq k_2\leq l-1$,
\item[(b)] 
$Q^{k_1}P_L^{-1}BF_2^{\dagger}\eta_2(n)$ and  $Q^{k_2}P_L^{-1}B F_2^{\dagger}\eta_2(n)$ 
are independent for any  $0\leq k_1\neq k_2\leq l-1$,
\item[(c)] $P_L^{-1}BF_2^{\dagger}\eta_2(n)$ and $F_2^{\dagger}\eta_2(n)$
           are independent,
\end{itemize}
then 
$\hat{\eta}_2^{(l-1)}(n)$ and $F_1^{\dagger}\eta_1(n)$ 
in (\ref{3.10}) are independent, and 
$P_L\hat{\eta}_1^{(l)}(n)$ and $F_2^{\dagger}\eta_2(n)$
in (\ref{3.11})  are independent.
\end{TT}

When the block length $L$ of the interleaver $P_L$ is sufficiently large,
such as $L>>K$,
and the interleaver $P_L$ is sufficiently random, the conditions 
(a)-(c) in Theorem 2 usually hold. 
By the forms of $Q^k$ with $Q$ defined in (\ref{3.13}),
in order to have the above independences (a)-(b), the operators $A$ and $B$
between $P_L$ and its inverse $P_L^{-1}$ should
not be $\alpha I_K$ for any nonzero constants $\alpha$.
As a conclusion, we have proved the following
corollary.

\begin{CC}
When the block length $L$ of the interleaver $P_L$ is sufficiently large 
and the interleaver $P_L$ is sufficiently random,
and the operators  $(I_K+F_1^{\dagger}F_1)$ and  $(I_K+F_2^{\dagger}F_2)$
are not equal to any nonzero constant multiples of
the identity operator $I_K$,
then $\hat{\eta}_2^{(l-1)}(n)$ and $F_1^{\dagger}\eta_1(n)$  
in (\ref{3.10}) are independent, and  
$P_L\hat{\eta}_1^{(l)}(n)$ and $F_2^{\dagger}\eta_2(n)$ 
in (\ref{3.11})  are independent.
\end{CC}

For an interleaver $P_L$, condition (c) usually holds. When the operator
$Q=P_L^{-1}BP_LA$ in (\ref{3.13}) maps a random process $X_1$ to another
random process $X_2$ so that $X_2$ is independent of $X_1$, then
conditions (a)-(b) also hold. This proves the following result.

\begin{CC}
For an interleaver $P_L$, if the operator $Q$ in (\ref{3.13})
maps a random process to an independent random process, then
$\hat{\eta}_2^{(l-1)}(n)$ and $F_1^{\dagger}\eta_1(n)$  
in (\ref{3.10}) are independent, and  
$P_L\hat{\eta}_1^{(l)}(n)$ and $F_2^{\dagger}\eta_2(n)$ 
in (\ref{3.11})  are independent.
\end{CC}

This result is potentially useful in designing a good interleaver.
Notice that in theory the interleaver can be replaced by any invertible 
decorrelator but the computational complexity has to be taken into account
when the block length $L$ is large.

In what follows, we assume that the conditions in Corollary 2 hold
and therefore the random variables 
$\hat{\eta}_2^{(l-1)}(n)$ and $F_1^{\dagger}\eta_1(n)$   
in (\ref{3.10}) are independent, and   
the random variables $P_L\hat{\eta}_1^{(l-1)}(n)$ 
and $F_2^{\dagger}\eta_2(n)$  
in (\ref{3.11})  are also independent. 
We next want to calculate the mean powers of the decoded noises
$\hat{\eta}_1^{(l)}(n)$ and $\hat{\eta}_2^{(l)}(n)$
in (\ref{3.10})-(\ref{3.11})
by utilizing the calculation (\ref{2.18}) in Section 2.

Let $|\alpha_k|^2$, $1\leq k\leq K$, and $|\beta_k|^2$, $1\leq k\leq K$,
be the eigenvalues of the matrices $F_1^{\dagger}F_1$
and $F_2^{\dagger}F_2$, respectively. By the optimality on codes 
$G_1$ and $G_2$ in Corollary 1, in what follows we assume condition
(\ref{2.24}) holds for both $\alpha_k$ and $\beta_k$, i.e., 
\begin{equation}\label{3.16} 
|\alpha_k|^2=\left\{ \begin{array}{ccc}  
1, & 1\leq k\leq N_1-K, &\mbox{when } N_1-K<K,\\ 
\frac{N_1-K}{K}, & 1\leq k\leq K, & \mbox{when } N_1-K\geq K,
\end{array}\right. 
\end{equation} 
and
\begin{equation}\label{3.17} 
|\beta_k|^2=\left\{ \begin{array}{ccc}  
1, & 1\leq k\leq N_2-K, &\mbox{when } N_2-K<K,\\ 
\frac{N_2-K}{K}, & 1\leq k\leq K, & \mbox{when } N_2-K\geq K.
\end{array}\right. 
\end{equation} 
Let $\sigma_{i,l}^2$ denote the mean power
of the random processes $\hat{\eta}_i^{(l)}(n)$ for $i=1,2$,
and $l=0,1,2,...$,
in (\ref{3.9})-(\ref{3.11}). 
Notice that the final decoded noise mean power in Fig. 2(b)
is $\sigma_{2,l}^2$, which is calculated in the following. 
Similar to what was studied in Section 2,
for the parameters $K,N_i$ for $i=1,2$ there are  four cases:
$0<N_1-K<K$ and $0<N_2-K<K$; $0<N_1-K<K$ and $0<N_2-K\geq K$;
$0<N_1-K\geq K$ and $0<N_2-K<K$; and $0<N_1-K\geq K$ and $0<N_2-K\geq K$.
In the last three cases, either one of  $I_K+F_i^{\dagger}F_i$ for $i=1,2$
or both of them are $\frac{N}{K}I_K$,  when their eigenvalues
$\alpha_k$ and $\beta_k$ take the forms in (\ref{3.16})-(\ref{3.17}).
Therefore, the last three cases do not satisfy Corollary 2.
This implies that, in these three cases, non-optimal parallel 
codes $G_1$ and $G_2$ need to be used for the independences
of the decoded noises in the iterations of the decoding. Also, 
in the last three cases, the turbo code rates are small, which
are usually not used in many applications. 
With these observations, we now consider the first case, i.e., 
$N_1-K<K$ and $N_2-K<K$. In this case,
the eigenvalues of matrices $F_i^{\dagger}F_i$ for $i=1,2$
are $1$ and $0$, and hence the matrices
$I+F_i^{\dagger}F_i$ for $i=1,2$ are not equal 
to $\alpha I_K$ for any constant $\alpha$, i.e.,
the conditions for the independences in Corollary 2 satisfy.

Similar to formula (\ref{2.18}) and by the 
independences of 
$\hat{\eta}_2^{(l-1)}(n)$ and $F_1^{\dagger}\eta_1(n)$, and 
$P_L\hat{\eta}_1^{(l-1)}(n)$ and $F_2^{\dagger}\eta_2(n)$  
in (\ref{3.10})-(\ref{3.11}), we have, for $l=1,2,...$,
and $\sigma_{1,0}^2=\sigma_{2,0}^2=\sigma_{\eta}^2$, 
\begin{eqnarray}
\sigma_{2,l}^2 & = & \frac{7K-3N_2}{4K}\sigma_{1,l}^2+
\frac{N_2-K}{4K}\sigma_{\eta}^2,\label{3.18}\\
\sigma_{1,l}^2 & = & \frac{7K-3N_1}{4K}\sigma_{2,l-1}^2+
\frac{N_1-K}{4K}\sigma_{\eta}^2. \label{3.19}
\end{eqnarray}
Thus,
\begin{eqnarray}
\sigma_{2,l}^2 & = & \frac{7K-3N_2}{4K} \frac{7K-3N_1}{4K}
\sigma_{2,l-1}^2 +\left[ \frac{7K-3N_2}{4K} \frac{N_1-K}{4K}
+  \frac{N_2-K}{4K}\right]\sigma_{\eta}^2 \nonumber\\
 & = & \frac{49K^2-21K(N_1+N_2)+9N_1N_2}{16K^2}\sigma_{2,l-1}^2
+\frac{7K(N_1+N_2)-3N_1N_2-11K^2}{16K^2}\sigma_{\eta}^2.
\label{3.20}
\end{eqnarray}
By the assumption $0<N_1-K, N_2-K<K$,
\begin{equation}\label{3.21}
0<\frac{7K-3N_2}{4K}<1,\,\,\,\,0<\frac{7K-3N_1}{4K}<1.
\end{equation}
Thus the recursive formula (\ref{3.20})  converges. Let $\sigma_2^2$
be the limit of $\sigma_{2,l}^2 $ when $l$ tends to infinity.
Then 
\begin{equation}\label{3.22}
\sigma_2^2=\lim_{l\rightarrow \infty}
\sigma_{2,l}^2=\frac{7K(N_1+N_2)-3N_1N_2-11K^2}
{21K(N_1+N_2)-9N_1N_2-33K^2}\sigma_{\eta}^2 =\frac{1}{3}\sigma_{\eta}^2.
\end{equation}
Surprisingly, the above limit is independent
of the rates $K/N_1$ and $K/N_2$. 
Notice that the recursive formula (\ref{3.20}) for $\sigma_{2,l}^2$
is symmetric with the parameters $N_1$ and $N_2$.
Hence, the same recursive formula holds
for $\sigma_{1,l}^2$. Thus the limit $\sigma_1^2$ of $\sigma_{1,l}^2$
when $l$ tends to infinity is also
\begin{equation}\label{3.23}
\sigma_1^2=\lim_{l\rightarrow \infty} \sigma_{1,l}^2= 
\frac{1}{3}\sigma_{\eta}^2.
\end{equation}

By (\ref{3.21}), the decoded noise mean power
$\sigma_{i,l}^2$ decreases when the iteration number $l$ increases
for both $i=1,2$ with the following decreasing rates: for $l=0,1,...$,
and $\sigma_{1,0}^2=\sigma_{2,0}^2=\sigma_{\eta}^2$,
\begin{eqnarray}
\sigma_{1,l}^2-\sigma_{i,l+1}^2 & = & \sigma_{\eta}^2
\frac{N_1-K}{2K}\Gamma^l, \label{3.24}\\
\sigma_{2,l}^2-\sigma_{2,l+1}^2 & = & \sigma_{\eta}^2
\frac{4K(N_1-K)+(N_2-K)(7K-3N_1)}{8K^2} \Gamma^l, \label{3.25}
\end{eqnarray}
where
\begin{equation}\label{3.26}
\Gamma = \frac{7K-3N_2}{4K} \cdot \frac{7K-3N_1}{4K}<1.
\end{equation}
Thus, from (\ref{3.24})-(\ref{3.25}), we have: for $l=1,2,...$,
\begin{eqnarray}
\sigma_{1,l}^2 & = & \sigma_{\eta}^2\left[ 1-\frac{N_1-K}{2K}\frac{1-\Gamma^l}{1-\Gamma}
\right], \label{3.27}\\
\sigma_{2,l}^2 & = & \sigma_{\eta}^2\left[ 1-\frac{4K(N_1-K)+(N_2-K)(7K-3N_1)}{8K^2}\frac{1-\Gamma^l}{1-\Gamma}
\right]. \label{3.28}
\end{eqnarray}

From (\ref{3.22}), one can see that  the coding gain is $3$.
Notice that, in this case, the overall rate for the turbo code
is $K/N=K/(N_1+N_2-K)>1/3$ when $N_1-K<K$ and $N_2-K<K$.
For a non-turbo code with rate $K/N$, 
by (\ref{2.26}) the optimal coding 
gain is less than $N/K<3$. This proves that
coding gain for turbo codes in the limit sense in this case is larger than 
any non-turbo codes with the same rate.
The gain of a turbo code over a non-turbo code with rate $K/N$ is
\begin{equation}\label{3.29}
\gamma_{turbo}(K,N)\stackrel{\Delta}{=}\frac{\sigma_{\eta}^2/
\sigma_{2}^2}{N/K}=\frac{3K}{N_1+N_2-K},\,\,\,
K<N_1<2K,\,\,K<N_2<2K,
\end{equation}
which is called the {\em turbo gain}.

In summary, the following theorem is proved.

\begin{TT}
Assume a turbo code with two parallel codes $G_1$ with rate $K/N_1$
and $G_2$ with rate $K/N_2$ as in (\ref{3.1}) such that 
$K<N_i<2K$ for $i=1,2$, and  (\ref{3.16})-(\ref{3.17}) hold.
Assume the conditions in Corollary 2 hold for the decorrelating 
property of the interleaver. 
Then the decoded noise mean power $\sigma_{2,l}^2$ decreases
with the decreasing rate (\ref{3.25})-(\ref{3.28}) as the iteration number $l$ increases.
The limit and the best decoded noise mean power 
are $\sigma_{\eta}^2/3$ with the original noise power $\sigma_{\eta}^2$,
which 
is independent of the 
rates $K/N_1$ and $K/N_2$, and 
can not be achieved by any non-turbo code with the same rate.
The turbo gain of turbo codes 
over non-turbo codes is formulated in (\ref{3.29}).
\end{TT}

Since the limit of the decoded noise power is independent of
the rates when $K<N_i<2K$ for $i=1,2$, the following 
corollary is clear.

\begin{CC}
The limit of the decoded noise mean power of the iterative decoding
of a turbo code with rate $K/(K+2)$ for any positive integer $K$
is $\sigma_{\eta}^2/3$, where $\sigma_{\eta}^2$
is the noise power before the decoding.
\end{CC}

Comparing to the conventional rate $K/(N_1+N_2-K)$ codes, by (\ref{2.25})
and (\ref{3.28}),
when the iteration number $l$ satisfies the following lower bound,
the decoded noise mean power for turbo codes with the same rate 
will be smaller than the ones for the conventional codes:
\begin{equation}\label{3.30}
l>\frac{\log\left(1-\frac{(1-\Gamma)4K(N_1+N_2-2K)}{4K(N_1-K)+(N_2-K)(7K-3N_1)}\right)}{\log \Gamma},
\end{equation}
where $\Gamma$ is defined in (\ref{3.26}).

\begin{CC}
When the number $l$ of iterations in turbo code decoding
satisfies the lower bound (\ref{3.30}), the decoded
noise mean power at the $l$th iteration is smaller than
the ones for the conventional codes with the same rate.
\end{CC}

An intuitive explanation for the above turbo gain 
is due to the iterations. 
Under the assumption of the independence
of the decoded noise after the long interleaving,
the new noise in the decoded signal $\hat{X}_2(n)$ in
Fig. 2(b) decreases when the number of
iterations increases, while the noises in the parity check 
parts $\tilde{Y}_i(n)$ for $i=1,2$ do not change
at each iteration of the decoding. This implies
that, the more parity check parts 
are,   the less turbo gain is. In other words, 
the turbo gain $\gamma_{turbo}(K,N)$ is 
a decreasing function of variable $N/K$. 
Since the independence assumption in Theorem 2 is an ideal assumption,
by Theorem 3, the following lower bound applies to the
decoded noise mean powers in the iterative decoding.

\begin{CC}
Let $\sigma_{2,l}^2$ be the noise mean power 
after the $l$th iterative decoding of a turbo code
with two parallel codes of rates $K/N_1$ and $K/N_2$,
respectively.
Then, when $K<N_1,N_2< 2K$, 
\begin{equation}\label{3.31}
\sigma_{2,l}^2\geq \frac{1}{3}\sigma_{\eta}^2,\,\,\,\mbox{ for }l\geq 0,
\end{equation}
where $\sigma_{\eta}^2$ is the original noise 
power before decoding.
\end{CC}

As an example, we consider two rate $2/3$ codes $G_1$ and $G_2$
in (\ref{3.1}), i.e., $K=2$ and $N_1=N_2=1$. 
Thus, the turbo code rate is $1/2$. By (\ref{2.25}) in Theorem 1, the decoded
noise mean power for the optimal rate $1/2$ non-turbo code 
is $\sigma_{\eta}^2/2$, while the limit of the decoded
noise mean power of this turbo code is $\sigma_{\eta}^2/3$.
We will see later from the numerical examples  in Section 4
that this limit is usually reached after only a few iteration
steps. 

The calculations (\ref{3.18})-(\ref{3.19}) 
of the noise mean powers $\sigma_{i,l}^2$
of $\hat{\eta}_i^{(l)}(n)$ for $i=1,2$ in (\ref{3.10})-(\ref{3.11})
are only for the optimal codes $G_1$ and $G_2$ with
the eigenvalues in (\ref{3.16})-(\ref{3.17}). The calculations for general
parallel codes can be done by using the formulas (\ref{2.14}) 
and (\ref{2.21}), where $|\alpha_k|$ may not be equal to each other. 
The limit of the decoded noise mean power can also be calculated.
The calculations are certainly more tedious than (\ref{3.18})-(\ref{3.19}). 
When non-optimal codes $G_1$ and $G_2$ are used, 
the operators $I_K+F_i^{\dagger}F_i$ for $i=1,2$
always satisfy the condition in Corollary 2 for the independence for any 
rates $K/N_i$, which do not necessarily 
satisfy $K<N_i<2K$, $i=1,2$, as in the above calculations.
For the other three cases, i.e., $K<N_1<2K, K<N_2\geq 2K$;
$K<N_1\geq 2K, K<N_2<2K$; and $K<N_1\geq 2K, K<N_2\geq 2K$,
similar calculations can be done by using the formulas (\ref{2.14})  
and (\ref{2.21}), when non-optimal parallel codes with rates 
$K/N_i$, $i=1,2$, are used. 
Although  turbo codes
with only two parallel codes have been studied, turbo codes with more parallel
codes can be similarly studied with more tedious formulas than (\ref{3.18})-(\ref{3.19}).
The interested readers can derive all of these extensions  by
themselves. We however omit the details here. 
In the next section, we want to see some numerical examples.

\section{Numerical Simulations}
\setcounter{equation}{0}

In this section, we consider two turbo codes defined on the complex field
with two different rates $1/2$ and $3/5$. In these two examples,
the noise power before decoding is chosen $1$, i.e., $\sigma_{\eta}^2=1$.

The first turbo code is: $N_1=N_2=1$ and $K=2$, in this case, the turbo 
code rate is $1/2$. The two parallel codes
are $G_1$ and $G_2$ with their parity check
matrices $F_1=F_2=(1/\sqrt{2}, -1/\sqrt{2})$
in (\ref{3.1}), respectively. In this case, 
the two eigenvalues of $F_i^{\dagger}F_i$ are $1$ and $0$, and 
$$
I_2+F_i^{\dagger}F_i=\left[\begin{array}{cc}
1.5 & -0.5 \\ -0.5 & 1.5 \end{array} \right],
$$
for $i=1,2$. Certainly, they satisfy the  condition in Corollary 2.
The interleaver $P_L$ is the block interleaver with length 
$200\times 100=20000$ with row vectors (linewise) writing in and column vectors
(columnwise) reading out, and $200$ by $100$ matrix size.
The theoretical curves for both 
$\sigma_{1,l}^2$ (after Decoder I) 
and $\sigma_{2,l}^2$ (after Decoder II)
in (\ref{3.18})-(\ref{3.19}) of the 
decoded noise mean powers vs. iteration number $l$ are plotted
in Fig. 3 with marks o and $\Box$, respectively.
The corresponding simulated curves are also plotted in
Fig. 3 with marks x and $+$, respectively. One can see that
the simulated and theoretical curves coincide with each other.
In Fig. 3, the lower bounds of the decoded noise mean powers
for both turbo codes and the conventional codes with rate 1/2
are also shown. 
In this example, the lower bound in (\ref{3.30})
on the number $l$ of the iterations
in turbo codes is $1.4748$, i.e., when $l\geq 2$, the decoded
noise mean power at the $l$th iteration is smaller than 
the ones for the conventional codes
(the curves when $l\geq 2$ is below the lower bound
$\sigma_{\eta}^2/2$ for rate $1/2$ non-turbo codes), which is precisely 
illustrated by the simulations shown in Fig. 3.

The second turbo code is: $N_1=N_2=1$ and $K=3$, in this case, the turbo 
code rate is $3/5$. The two parallel codes
are $G_1$ and $G_2$ with their parity
check matrices $F_1=F_2=(1/\sqrt{3}, 1/\sqrt{3},-1/\sqrt{3} )$
in (\ref{3.1}), respectively. In this case, 
the three eigenvalues of $F_i^{\dagger}F_i$ are $1$, $0$ and $0$
for $i=1,2$.
Similar to the first example, these codes also satisfy the condition
in Corollary 2. The interleaver in this turbo code has length $600\times 50=30000$ and 
has the same type as the one in the first example. 
The theoretical and simulated curves of the decoded noise mean powers
vs. the iteration number are plotted in Fig. 4. 
Similarly, the theoretical and simulated results coincide. 
Also, these curves converges to the same lower bound as the one
in the first example, that is $\sigma_{\eta}^2/3$. 
This illustrates that the decoded noise mean power of 
turbo codes are eventually independent of the rates 
when the rates of the two parallel codes are less than $1/2$.
However, one can clearly see from these two examples that 
the convergence speeds with different rates are different. 
In this example, 
 the lower bound in (\ref{3.30})
on the number $l$ of the iterations
in turbo codes is $1.2047$, i.e., when $l\geq 2$, the decoded
noise mean power at the $l$th iteration is smaller than 
the ones for the conventional codes, which is also precisely 
illustrated by the simulation shown in Fig. 4.
Our numerous examples show that all the interleavers
with square matrices do not perform well
and actually the iterations may increase the decoded 
noise mean power,  and the performance 
is sensitive to the choice of interleavers.

\section{Conclusion}

In this paper, we have analytically shown that 
the decoded noise mean power for turbo codes is smaller 
than the one for the conventional codes by  extending
them from finite fields to the complex 
field. It has been shown that the decoded 
noise mean power decreases when the iteration number increases
in the iterative turbo code decoding, where
the  analytic decreasing rate has been given.
We have also provided the limit of the 
decoded noise mean power in the iterative decoding.
The limit  is one third of the original noise power
before the decoding, when the two parallel codes 
in turbo codes have rates less than $1/2$,
which can not be achieved by any non-turbo codes
with the same rate. All these results are built upon 
the assumption on the interleavers that decorrelates
the decoded noises. From the results in this paper,
the role of the interleavers is very clear.
It turns out that the condition for the interleaver
$P_L$ is that the operator $P_L^{-1}(I_K+F_2^{\dagger}F_2)^{-1}
P_L(I_K+F_1^{\dagger}F_1)^{-1}$ maps a random process
to another independent random process, where 
$(I_K+F_i^{\dagger}F_i)^{-1}$ for $i=1,2$ are 
corresponding to the two cascaded decoders. 
The theoretical analysis has been illustrated
by two numerical examples with rates 1/2 and 3/5.
Our simulation results show that,
after a few iteration decoding steps,
the turbo code performance becomes better than 
the conventional code (non-turbo code) performance, and
also the performance strongly depends on the choice
of the interleavers.

Although the discussions in this paper are
for codes defined on the complex field
and  from the digital signal processing point of view,
it is believed that a similar analysis applies to
codes defined on finite fields and even possibly a better
performance can be achieved 
by taking the signal constellation and quantization into account.
Further researches along this direction 
are under our current investigations. 

This paper was written in the June of 1997. 

\section*{References}

\noindent
[1] C. Berrou, A. Glavieux, and P. Thitimajshima, ``Near 
Shannon limit error-correcting coding and decoding: turbo
codes,'' {\em Proc. ICC'93}, pp. 1064-1070, Geneve, Switzerland,
May 1993.

\noindent
[2] D. Divsalar and F. F. Pollara, ``Turbo codes
for PCS applications,'' {\em Proc. ICC'93}, Seattle, WA, June 1995.

\noindent
[3] P. Jung and M. Nasshan, ``Performance evaluation
of turbo codes for short frame transmission
systems,'' {\em Electron. Lett.}, vol. 30, 
pp. 111-113, Jan. 1994.

\noindent
[4] P. Jung and M. Nasshan, ``Dependence of the error
performance of turbo codes on the interleaver structure
in short frame transmission
systems,'' {\em Electron. Lett.}, vol. 30, 
pp. 287-288, Feb. 1994.

\noindent
[5] S. Benedetto and G. Montorsi, ``Average
performance of parallel concatenated block codes,''
{\em Electron. Lett.}, vol. 31, pp. 156-158, Feb. 1995.

\noindent
[6] S. Benedetto and G. Montorsi, ``Performance evaluation
of turbo-codes,'' {\em Electron. Lett.}, vol. 31, pp. 163-165, Feb. 1995.

\noindent
[7] S. Benedetto and G. Montorsi, ``Unveiling
turbo codes: some results on parallel concatenated coding
schemes,'' {\em IEEE Trans. on Information Theory},
vol. 42, pp. 409-428, March 1996.

\noindent
[8] J. Hagenauer, E. Offer, and L. Papke, ``Iterative
decoding of binary block and convolutional codes,''
{\em IEEE Trans. on Information Theory},
vol. 42, pp. 429-445, March 1996.

\noindent
[9] C. Berrou and A. Glavieux, ``Near optimum error correcting
coding and decoding: turbo codes,'' {\em IEEE Trans. on Communications},
vol. 44, 1996.

\noindent
[10] X.-G. Xia, ``On modulated coding and least square decoding
via coded modulation and Viterbi decoding,'' 
technical report \#97-6-2,
Department of Electrical and Computer 
Engineering, University of Delaware, 1997.

\noindent
[11] H. Liu and X.-G. Xia, ``Precoding techniques for undersampled
multi-receiver communication systems,'' technical report \#97-3-1,
Department of Electrical Engineering, University of Delaware, 1997.

\noindent
[12] X.-G. Xia and H. Liu,  ``Polynomial ambiguity resistant precoders: theory and applications in ISI/multipath  cancellation,''
technical report \#97-5-1,
Department of Electrical Engineering, University of Delaware, 1997.

\noindent
[13] X.-G. Xia and G. Zhou, ``On optimal ambiguity resistant precoders in ISI/multipath cancellation,'' technical report \#97-5-2,
Department of Electrical Engineering, University of Delaware, 1997.

\noindent
[14] Z. Xie, R. T. Short, and C. K. Rushforth, ``A family
of suboptimum detectors for coherent multiuser
communications,''
{\em IEEE J. Select. Areas Communications}, vol. 8,
pp. 683-690, May 1990.

\noindent
[15] U. Madhow and M. Honig, ``MMSE interference
suppression for direct-sequence spread spectrum
CDMA,'' {\em IEEE Trans. on Communications}, 
vol. 42, pp. 3178-3188, Dec. 1994.

\noindent
[16] H. V. Poor and S. Verd\'{u}, ``Probability of error
in MMSE multiuser detection,'' {\em IEEE Trans. on Information
Theory}, vol. 43, pp. 858-871, May 1997.

\noindent
[17] A. J. Viterbi, {\em CDMA : Principles of Spread Spectrum Communication},
Reading, Mass. : Addison-Wesley Pub. Co., 1995.

\noindent
[18] IEEE Information Theory Society Newsletter,
 ``A conversation with G. David Forney, Jr.,'' vol. 47, No. 2, June 1997.

\clearpage

\begin{figure}[htbp]
 \begin{center}
 \epsfysize=5.4in
 \leavevmode\epsffile[62   201   598   616 ]{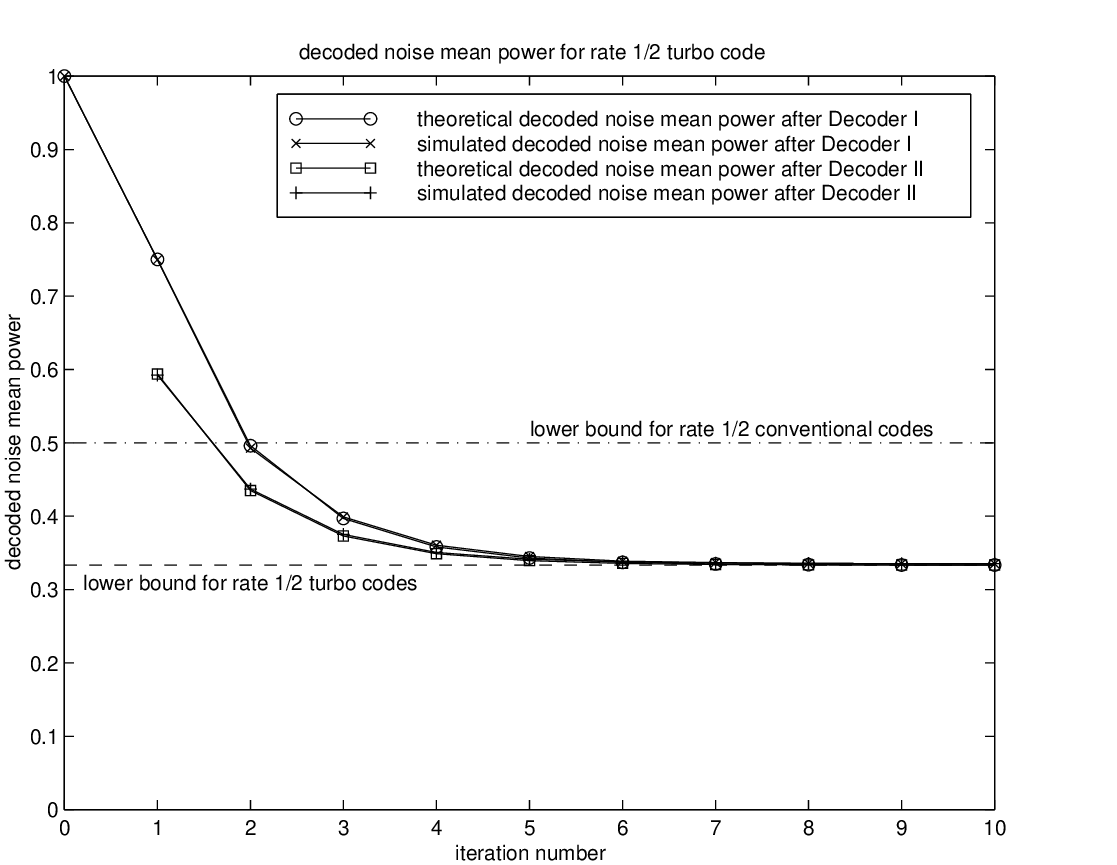}
\end{center}
\caption{Decoded noise  mean powers vs. iteration numbers for
the rate 1/2 turbo code}
\end{figure}

\clearpage

\begin{figure}[htbp]
 \begin{center}
 \epsfysize=5.4in
 \leavevmode\epsffile[62   201   598   616 ]{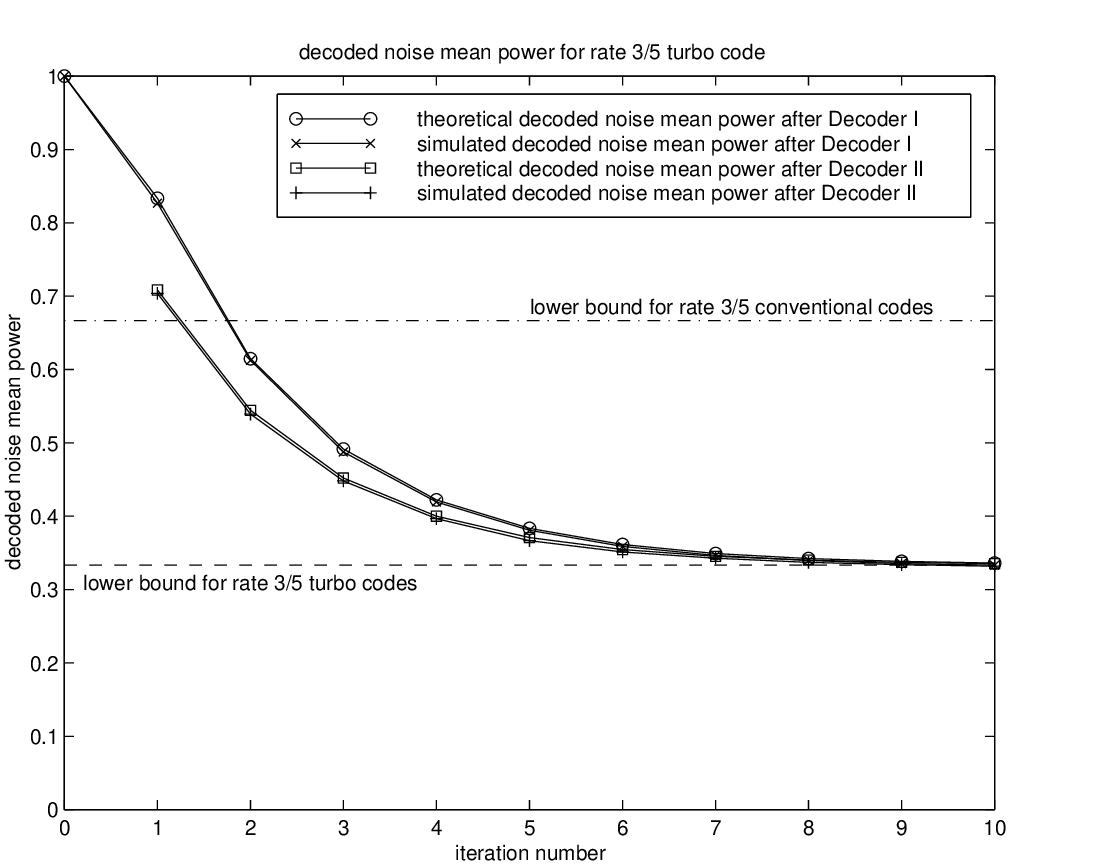}
\end{center}
\caption{Decoded noise  mean powers vs. iteration numbers for
the rate 3/5 turbo code}
\end{figure}

\end{document}